\documentclass[%
reprint,
amsmath,amssymb,
aps,
physrev,
]{revtex4-2}

\usepackage[OT1]{fontenc}
\usepackage{lmodern}
\usepackage{sansmathfonts}

\usepackage{graphicx}
\usepackage{dcolumn}
\usepackage{bm}
\usepackage{hyperref}
\usepackage{geometry}

\usepackage{siunitx}
\usepackage{ulem}
\usepackage{nccmath}
\usepackage{cancel}

\begin{document}

\title{
Reflection of forward volume exchange spin waves at ferromagnetic junctions interpreted via continuity of angular momentum density flux
}

\author{Luis Sánchez-Tejerina}
 \email{luis.sanchez-tejerina@uva.es}
 \affiliation{Department of Electricity and Electronics, University of Valladolid, Valladolid, Spain.}%
\author{Luis M. Moreno-Ramírez}%
\affiliation{Department of Electricity and Electronics, University of Valladolid, Valladolid, Spain.}%
\author{Óscar Alejos Ducal}%
\affiliation{Department of Electricity and Electronics, University of Valladolid, Valladolid, Spain.}%

\date{\today}%

\begin{abstract}
Magnonics investigates the dynamic excitations of magnetically ordered media, known as spin waves, and their application in functionalized devices. To efficiently control spin wave based devices minimizing losses and interferences from unwanted reflections, a clear picture of the scattering mechanism is needed. Here, we focus on the reflection of spin waves produced at the interface between two ferromagnetic media. We show that it is possible to understand the reflection mechanism based on the continuity of the angular momentum density flux. This constraint allows us to analytically compute the reflection coefficient and the stationary wave envelope in the first medium, as well as the damped propagation in the second. In addition, the analytical expressions are compared with full micromagnetic simulations, showing excellent agreement.
\end{abstract}


\maketitle

A magnetically ordered sample can be excited by different mechanisms, including microwave~\cite{Schneider:2008,Demidov:2009b,Demidov:2009,Vogt:2014,Wang:2020,Connelly:2021,Levchenko:2026} and optical~\cite{Kimel:2005,Kirilyuk:2010,Schmidt:2010,Dewhurst:2018,Sánchez-Tejerina:2023} electromagnetic fields or electrical currents~\cite{Demidov:2010,Madami:2011,Bechberger:2026}. These excitations, or spin waves (SWs), can be understood as a sum of elementary excitations described as harmonic waves, whose quantization leads to the concept of a magnon~\cite{Chumak:2015,Wang:2020,Serga:2010}. Magnons transport energy and momentum, thereby providing a promising platform to transmit, store, and process information. Moreover, the transmission of magnon currents in magnetic insulators occurs without associated Joule heating losses, allegedly allowing for the establishment of an energy-efficient alternative to traditional electronics for future computing devices~\cite{Serga:2010,Zheng:2022,Flebus:2024}.

\begin{figure*}
\includegraphics[width=0.85\textwidth]{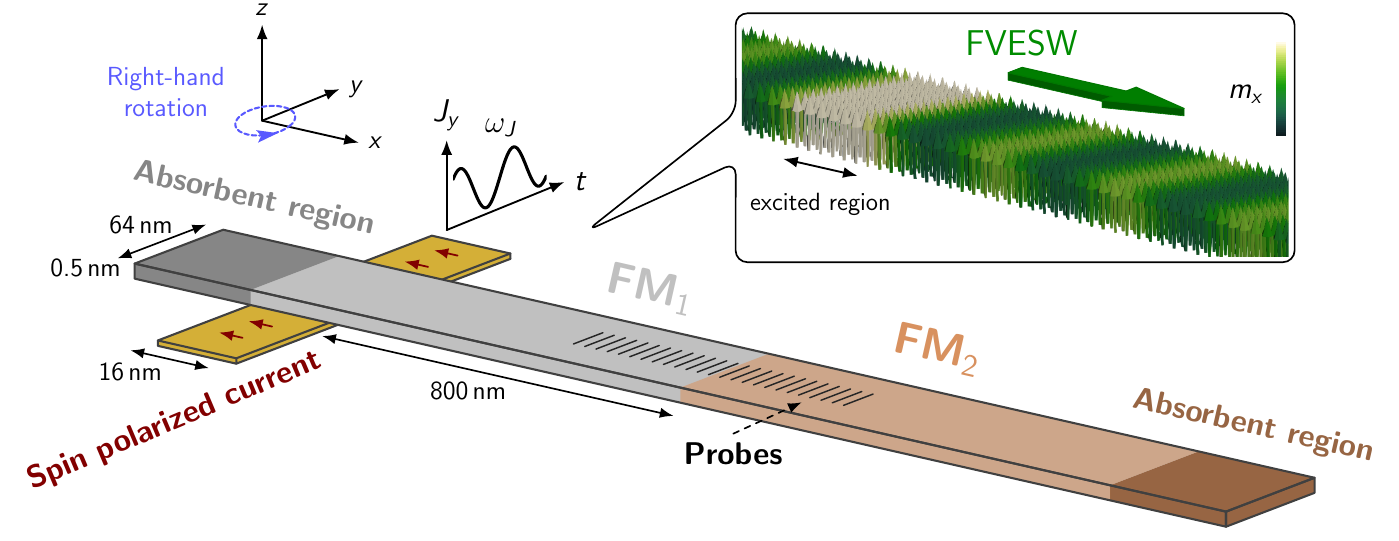}
\caption{\label{fig:esquema} \textbf{Schematic representation of the simulated configuration}. It consists of: (1) an excitation region, in which forward volume exchange spin waves (FVESWs) are generated in a ferromagnetic medium ($\text{FM}_{1}$) by a spin-polarized current, (2) absorbent regions positioned at the terminal portions of the strip to avoid undesired reflections, (3) a propagation region composed of $\text{FM}_{1}$ medium and an interface with a second ferromagnetic medium ($\text{FM}_{2}$) located at a distance of $\qty{800}{\nano\meter}$ from the excitation boundary. Probes are strategically positioned in the vicinity of the interface to record the associated scattering phenomena. The main dimensional parameters are specified within the illustration. A magnified view of the generated FVESW is presented in the inset. The reference system used in this study, as well as the right-hand rotation direction, are also included.}
\end{figure*}
A relevant aspect of SW propagation is the reflection and transmission through an interface. In general, such processes are governed by the conservation of certain physical quantities which impose specific boundary conditions at the interface. This problem has been studied for different scenarios, both theoretically~\cite{Gorobets:1998,Reshetnyak:2004,Alwahsh:2011,Kruglyak:2014,Poimanov:2019,Verba:2020} and numerically~\cite{Gruszecki:2015,Mieszczak:2020,Xing:2022}, where slightly different boundary conditions are considered, preventing a clear interpretation of the reflection phenomenon. In this study, we investigate the reflection of forward volume exchange spin waves (FVESWs) at an interface between two ferromagnetic (FM) media. We define an angular momentum density flux for circularly polarized modes, which allows us to derive analytical expressions for the corresponding reflection and transmission coefficients by imposing the continuity of this quantity. These expressions are compared with full micromagnetic simulations, obtaining excellent agreement.

To describe the magnetization dynamics of a FM sample, magnetization is treated as a continuum vector field $\vec{M}\left(\vec{r},t\right)$, defined for each position and time. Its evolution is described by the Landau-Lifshitz-Gilbert (LLG) equation~\cite{Sanchez-Tejerina:2025}

\begin{equation}\label{eq:llg_plain}
\frac{\mathrm{d}\vec{m}}{\mathrm{d} t} = -\gamma \vec{m}\times\vec{B}_{\mathrm{eff}} + \alpha \vec{m}\times\frac{\mathrm{d}\vec{m}}{\mathrm{d}t},
\end{equation}
where $\vec{m}=\frac{\vec{M}}{M_{\mathrm{s}}}$ is the unit vector along the magnetization direction, $M_{\mathrm{s}}$ the saturation magnetization, $\vec{B}_{\mathrm{eff}}$ the effective field including the exchange, anisotropy, and demagnetizing field, $\gamma$ the electron gyromagnetic ratio, and $\alpha$ the Gilbert damping. $\vec{B}_{\mathrm{eff}}$ is computed as the variational derivative of the total energy density. Assuming uniaxial anisotropy along $z$ (see Fig.~\ref{fig:esquema}) and thin film approximation (see \textit{supplementary material S1} for more information), the effective field would read
\begin{equation}\label{eq:Beff}
\vec{B}_{\mathrm{eff}} = \frac{2A}{M_{\mathrm{s}}}\nabla^2\vec{m} + \frac{2K_{\mathrm{eff}}}{M_{\mathrm{s}}} m_z\hat{u}_z ,
\end{equation}
where $A$ is the exchange stiffness and $K_{\mathrm{eff}}$ is the effective anisotropy. The analytical expressions derived are compared with the results of micromagnetic simulations obtained using the open-source solver MuMax³~\cite{Vansteenkiste:14}. The different physical quantities derived from micromagnetic simulations are labeled by the (sim) superscript. The material parameters considered here are of the order that can be found in CoFeB thin films with a MgO capping layer~\cite{Ikeda:2010,Hamalainen:2018,Kim:2022}. Although we performed a parametric study, we will take the following parameters as reference (indicated by $0$ subscript): $A_0=\qty{7.7}{\pico\joule\per\m}$, $M_{\mathrm{s},0}=\qty{1.71}{\mega\ampere\per\m}$, surface uniaxial anisotropy $K_{\mathrm{u},0}=K_{\mathrm{eff},0}+\frac{1}{2}\mu_0 N_z M^2_{\mathrm{s},0}=\qty{2.00}{\mega\joule\per\m\cubed}$, being $\mu_0$ the vacuum magnetic permeability and $N_z$ the Aharoni demagnetizing factor~\cite{Aharoni:98} for the out-of-plane direction, and Gilbert damping $\alpha_0=10^{-3}$. Our system is sketched in Fig.~\ref{fig:esquema}, where a thin uniaxial out-of-plane FM slab with thickness $d=\qty{0.5}{\nm}$, width $w=\qty{64}{\nm}$, and total length $\sim \qty{11}{\um}$ is considered. At both ends, regions with geometrically increasing damping are defined as absorbent regions to attenuate the SWs before they reach the end. The excitation region is located adjacent to the left absorbent region. Its width and length are fixed to $w_\mathrm{e}=\qty{64}{\nm}$ and $x_\mathrm{e}=\qty{16}{\nm}$, respectively. The system is assumed to be excited by an electric current of frequency $\omega_J$ through the spin Hall effect (SHE). Therefore, the LLG Eq.~\ref{eq:llg_plain} must be complemented with the torque due to the SHE~\cite{Emori:14,Sanchez-Tejerina:20}
\begin{equation}\label{eq:SHE_torque}
\vec{\tau}_{\mathrm{SH}} = \frac{\tau_{\mathrm{SH}}e^{-i\left(\omega_{J} t + \varphi\right)}}{S_{\mathrm{s}}} \vec{m}\times\left(\vec{m} \times \vec{\sigma}_{\mathrm{SH}} \right) ,
\end{equation}
\noindent where $\tau_{\mathrm{SH}}=\frac{\hbar \theta_{\mathrm{SH}}J_{\mathrm{e}}}{2|e|d}$, $\hbar$ being the reduced Planck constant, $\theta_{\mathrm{_{SH}}}=0.15$ the considered spin Hall angle, $J_{e}$ the amplitude of the electrical current density, $|e|$ the elementary charge, $\varphi$ the phase shift with respect to dynamic magnetization, $\vec{\sigma}_{\mathrm{SH}}$ the direction of the injected angular momentum density and $S_{\mathrm{s}}=\frac{M_{\mathrm{s}}}{\gamma}$ the saturation angular momentum density. A narrow strip was selected to ensure single-mode operation. SWs can be treated for small excitation amplitudes as a small deviation, $\delta\vec{m}$, from the equilibrium magnetization $\vec{m}_{\mathrm{eq}}$, where $\vec{m}=\vec{m}_{\mathrm{eq}}+\delta\vec{m}\,\mathrm{e}^{-i\left(\omega t - k_x x\right)}$. Assuming a negligible in-plane anisotropy, the solution for FVESW modes is a right-handed circularly polarized wave~\cite{Gurevich:1998}. Fig.~\ref{fig:esquema} sketches such a polarization. Solving Eq.~\ref{eq:llg_plain} under these assumptions yields (see \textit{supplementary material S2}) the dispersion relation for the first $k$-dependent FVESW mode

\begin{equation}\label{eq:disperssion}
\begin{split}
\omega =\frac{1-i\alpha}{1+\alpha^2}\frac{2\left(Ak_x^2+K_\mathrm{eff}\right)}{S_{\mathrm{s}}}= \omega' + i \omega'',
\end{split}
\end{equation}

\noindent $\omega'$ and $\omega''$ being real quantities. The group velocity is the derivative of the frequency $\omega '$ with respect to the wave number $\vec{k}$, i.e., reading

\begin{equation}\label{eq:vgr}
\begin{split}
\vec{v}_{\mathrm{gr}} = \frac{\partial \omega'}{\partial \vec{k}}=\frac{4Ak_x}{\left(1+\alpha^2\right)S_{\mathrm{s}}}\hat{u}_x.
\end{split}
\end{equation}

\noindent The amplitude is computed from the augmented LLG equation that includes the SHE torque (Eq.~\ref{eq:SHE_torque}). The procedure is detailed in \textit{supplementary material S2}. Assuming $\omega'=\omega_{J}$, the amplitude at the edge of the excitation region is
\begin{equation}\label{eq:amplitude}
\delta m_{\mathrm{e}} = \frac{\omega' x_{\mathrm{e}}}{4v_{\mathrm{gr}}} \frac{\tau_{\mathrm{SH}}\:\sin{\varphi}}{K_\mathbf{eff} + Ak^2_x} \mathrm{sinc} \left( k_x \frac{x_{\mathrm{e}}}{2}\right).
\end{equation}
This initial amplitude decays exponentially along the strip as the SWs propagate due to the term $e^{\omega''t}$. This term can be cast into a spatial dependence through the group velocity, which yields $e^{\alpha_{\mathrm{gr}} x}$, being the spatial attenuation coefficient
\begin{equation}\label{eq:attenuation}
  \alpha_{\mathrm{gr}}=\omega''/|\vec{v}_{\mathrm{gr}}|=-\alpha\frac{Ak_x^2+K_\mathrm{eff}}{2Ak_x}.
\end{equation} 
The phase shift between excitation and signal is $\varphi=\arctan\left(-1/\alpha\right)$. 

\begin{figure}
\centering
\includegraphics[width=\columnwidth]{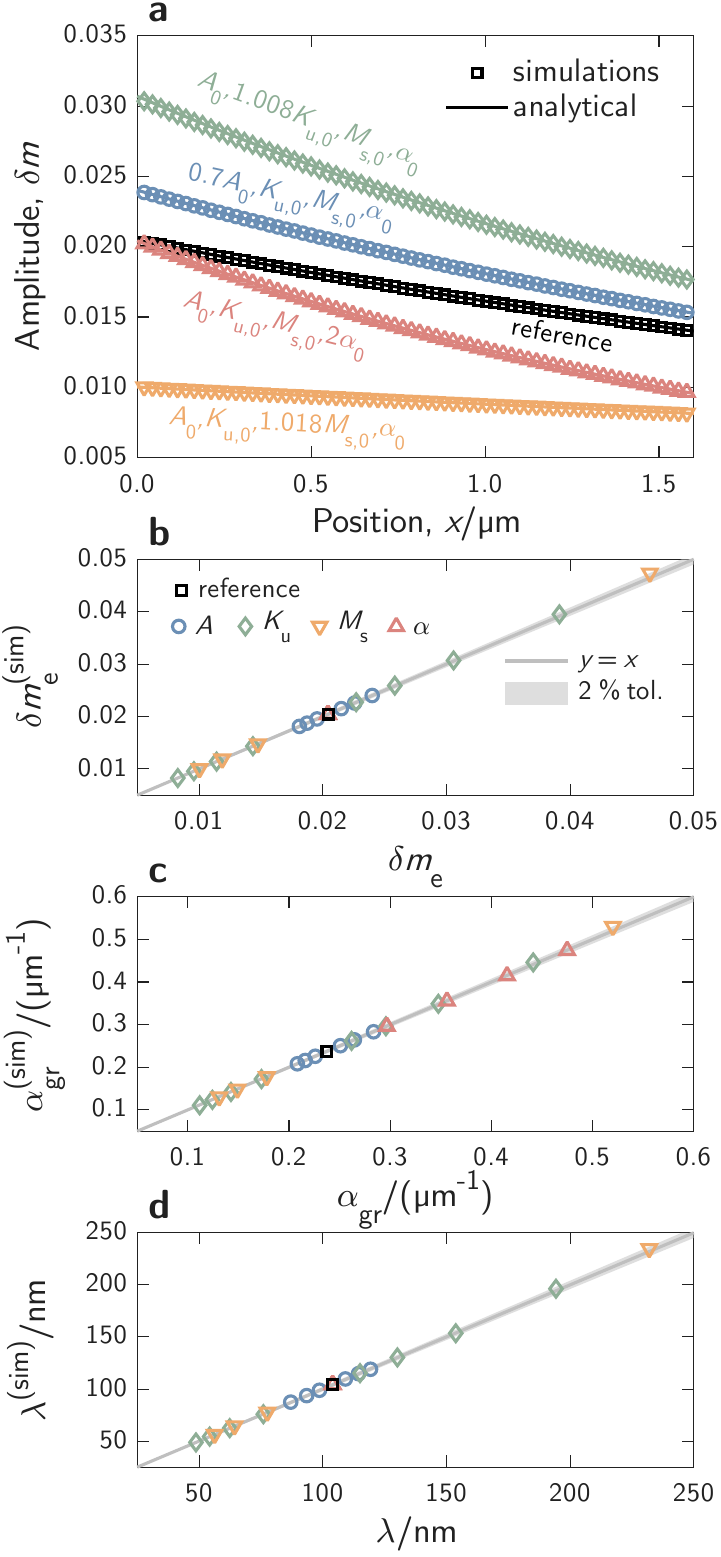}
\caption{\label{fig:1mat} \textbf{FVESWs in a ferromagnet}. (a) Amplitude dependence on the relative position to the excitation edge for various material parameters. The reference values $A_0$, $K_{\mathrm{u},0}$, $M_{s,0}$, and $\alpha_0$ are those indicated in the text. Solid lines represent analytical solutions while symbols denote micromagnetic simulations results. Comparison of the simulation results with the corresponding analytical solutions for (b) the amplitude at the excitation edge, (c) the spatial attenuation, and (d) the wavelength under different parameter modifications. Panels (b)–(d), include an identity line and a tolerance band of $\qty{2}{\percent}$.}
\end{figure}

Before the study on an FM junction, we validated the ability of our analytical model to predict the propagation characteristics of FVESW in our system. To do that, we performed micromagnetic simulations considering that both ferromagnetic materials $\text{FM}_{1}$ and $\text{FM}_{2}$ are the same. The system is excited at a frequency of $\omega_{J}/(2\pi)=\qty{7.20}{\GHz}$ and we considered varying magnetic parameters. $A$ ranges from $\qty{5.4}{\pico\joule\per\m}$ to $\qty{10.0}{\pico\joule\per\m}$, $\alpha$ from $10^{-3}$ to $2\cdot10^{-3}$, $M_{\mathrm{s}}$ takes values between $\qty{1.70}{\mega\ampere\per\m}$ and $\qty{1.74}{\mega\ampere\per\m}$, and $K_{\mathrm{u}}$ between $\qty{1.90}{\mega\joule\per\m\cubed}$ and $\qty{2.02}{\mega\joule\per\m\cubed}$. These parameter values were selected to produce substantial variability in the wave characteristics, while ensuring that the excitation frequency is sufficiently larger than the cut-off frequency. This condition avoids the excitation of undesired modes; see \textit{supplementary material S3} for more information. Fig.~\ref{fig:1mat}(a) shows the dependence on the position on the slab of the SW amplitude extracted from the full micromagnetic simulations (symbols) and the analytical formula (lines). The origin of positions is located at the interface between the excitation region and the non-forced region. Both micromagnetic and analytical results show the same amplitude, including the expected decay along the strip, demonstrating excellent agreement between both approaches. To further explore the observed agreement, we represent the amplitude at the origin, Fig.~\ref{fig:1mat}(b), the spatial attenuation coefficient $\alpha_{\mathrm{gr}}$, Fig.~\ref{fig:1mat}(c), and the wavelength $\lambda=2\pi/k_x$, Fig.~\ref{fig:1mat}(d) obtained from micromagnetic simulations versus analytical predictions (Eqs. \eqref{eq:amplitude}, \eqref{eq:attenuation}, and \eqref{eq:disperssion}). All points are observed to lie on the identity line with deviations below $\qty{2}{\percent}$. These results highlight the ability of the analytical model to describe the propagation of FVESW under the assumptions considered.

Once the analytical model is verified, we examined the behavior of the SWs when reaching the interface that separates the two magnetic media, $\text{FM}_{1}$ and $\text{FM}_{2}$. We assumed that both the direction of the angular momentum density and the flux of the angular momentum density are conserved, the latter defined as the magnitude of the angular momentum multiplied by the group velocity, that is, $\vec{J}_{S,z}=S_{\mathrm{s}}\left(\delta m\right)^{2}\vec{v}_{\mathrm{gr}}$. These assumptions lead, as described in the \textit{supplementary material S2}, to the following conditions

\begin{subequations}
\begin{align}
\frac{1-R}{1+R}=\frac{S_{\mathrm{s},2}v_{\mathrm{gr},2}}{S_{\mathrm{s},1}v_{\mathrm{gr},1}},\label{eq:reflection}\\
T= \frac{2S_{\mathrm{s},1}v_{\mathrm{gr},1}}{S_{\mathrm{s},1}v_{\mathrm{gr},1}+S_{\mathrm{s},2}v_{\mathrm{gr},2}}\label{eq:transmission},
\end{align}
\end{subequations}
\noindent where $R=\delta m^{\mathrm{r}}/\delta m^{\mathrm{i}}$ and $T=\delta m^{\mathrm{t}}/\delta m^{\mathrm{i}}$ are the reflection and transmission coefficients and $\delta m^{\mathrm{i}}$, $\delta m^{\mathrm{r}}$, and $\delta m^{\mathrm{t}}$ are the incident, reflected, and transmitted amplitudes, respectively. In this case, interference between incident and reflected waves in the first medium produces a normalized stationary wave pattern $p\left(x\right)=\frac{\delta m \left(x\right)}{\delta m^{\mathrm{i}}\left(0\right)}$ of the form

\begin{equation}\label{eq:stationary_wave}
p=\sqrt{e^{2\alpha_{\mathrm{gr,1}}x} + \lvert R\rvert ^{2}e^{-2 \alpha_{\mathrm{gr,1}}x} +\lvert R\rvert \cos{(2k_{x,1}x+\phi)}},
\end{equation}
where the origin is located at the interface and $\lvert R \rvert$ and $\phi$ are the modulus and the phase of the reflection coefficient, respectively.

\begin{figure*}
\centering
\includegraphics[width=\textwidth]{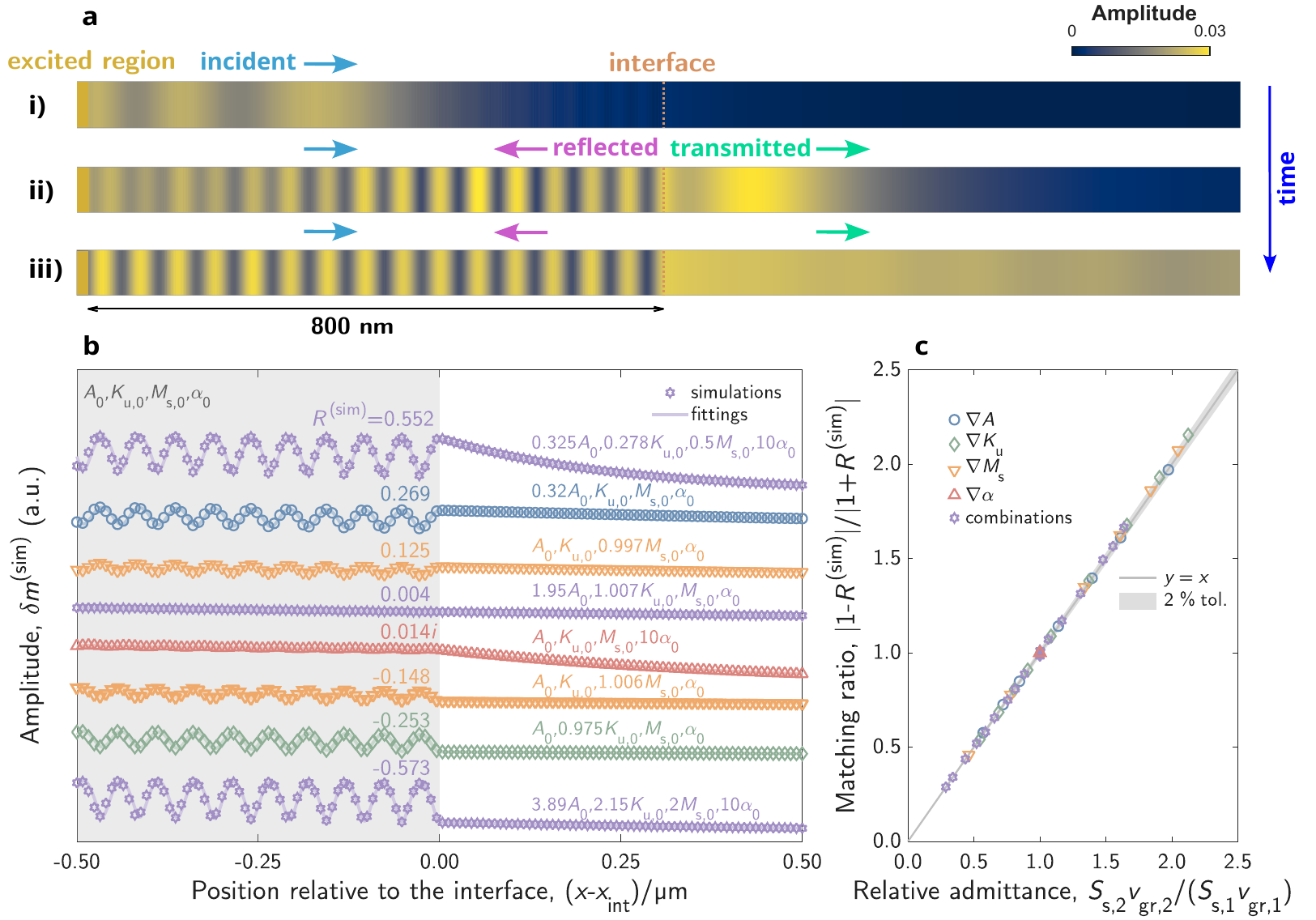}
\caption{\label{fig:2mat} \textbf{FVESWs in the vicinity of a ferromagnetic junction}. (a) Micromagnetic results of the spatial profile of the wave amplitude along the strip at selected time snapshots, illustrating the reflection and transmission phenomena and the subsequent formation of a standing wave in the first medium. (b) Simulation results (symbols) of the stationary wave amplitude in the vicinity of the interface for various parameters discontinuities. Corresponding fitted profiles (solid lines) to Eq. \ref{eq:stationary_wave}, yielding the respective reflection coefficients $R^{\text{(sim)}}$ as indicated. (c) Derived matching ratio between transmitted and reflected waves as a function of the relative admittance of the media (analytically calculated) for various parameter gradients at the interface. An identity line and a tolerance band of $\qty{2}{\percent}$ are included to highlight the excellent identity relation between both physical quantities.}
\end{figure*}

Figure~\ref{fig:2mat}(a) depicts three snapshots of micromagnetic simulations showing the amplitude of the FVESW at three different stages: i) during the first transitory when only an incident SW is present, ii) during the second transitory, when the SW has been reflected and transmitted through the interface between the two media, and iii) at the stationary state. %
The amplitude in the stationary state is shown in Fig.~\ref{fig:2mat}(b) where symbols correspond to full micromagnetic simulations while lines correspond to the fitting to Eq.~\ref{eq:stationary_wave}. From these fittings, the reflection coefficient,  $R^{\text{(sim)}}$, has been extracted. It is important to note that, in most cases, the phase is either $0$ or $\pi$, making the reflection coefficient real-valued and either positive or negative. The comparison with the analytical model (see Eq. \eqref{eq:reflection}) can be observed in Fig.~\ref{fig:2mat}(c) where the matching ratio, $\lvert 1-R^{\text{(sim)}} \rvert / \lvert 1+R^{\text{(sim)}} \rvert$, is represented using the reflection coefficient extracted from full micromagnetic simulations versus the relative admittance, $S_{\mathrm{s},2}v_{\mathrm{gr},2}/(S_{\mathrm{s},1}v_{\mathrm{gr},1})$, computed from the material parameters. The various colors and symbols represent changes in different parameters, except for the blue hexagon, which corresponds to a change in two or more parameters. Again, the agreement between both approaches is remarkable, with all points lying on the identity line within a tolerance of approximately $\qty{2}{\percent}$. Moreover, we are able to match two FM materials with a different set of parameters, as shown in the central envelope of Fig.~\ref{fig:2mat}(b). A reflection coefficient as small as $R^{\text{(sim)}}=\num{0.004}$ is computed even when both the exchange and anisotropy differ. Indeed, we attribute the small reflection to numerical errors rather than to the approximations of the model. A comment should be made on the case where solely the damping parameter is changed. As derived from the above, a negligible influence on the reflection coefficient due to a small change in group velocity should be expected from the analytical model. Indeed, the four cases analyzed (up to damping ratios of 1 to 10 at the interface) are superimposed in Fig.~\ref{fig:2mat}(c) at the point $\text{(1.0, 1.0)}$. However, the maximum magnitude of the reflection coefficient obtained from the fitting, although small, is not negligible ($\num{0.014}$). Moreover, we notice that the reflection coefficient is purely imaginary ($\phi^{\text{(sim)}}=\pi/2$), in contrast with all other cases where it is always real. This reactive behavior of the interface has been related to localized evanescent modes due to polarization mismatch~\cite{Poimanov:2019,Verba:2020}. More research is needed to clarify this point. In any case, this effect is marginal and would be veiled by any admittance or polarization mismatch.

In summary, we proposed boundary conditions based on a conservation argument to facilitate the analysis of the reflection and transmission of FVESW at an interface between two ferromagnets. In addition, we benchmarked the analytical results against numerical simulations, obtaining remarkable agreement. Although similar studies have been reported, the derived analytical expressions for the reflection and transmission coefficients vary, primarily due to the use of different boundary conditions. Furthermore, not all studies are systematically corroborated using micromagnetic simulations. Here, we aim to provide a clearer physical picture of the boundary conditions to be applied. These are found to be equivalent to the `natural boundary conditions' given by~\cite{Kruglyak:2014} but differ from other more recent studies~\cite{Poimanov:2019,Verba:2020,Gruszecki:2015,Xing:2022}. This approach also endorses the choice of the angular momentum density as a natural descriptor for the propagation of SWs. Future work will extend the angular momentum flux conservation condition to elliptically polarized SWs.

\begin{acknowledgments}
We acknowledge funding from Project PID2023-150853NB-C31 funded by MICIU/AEI/10.13039/501100011033 and FEDER, UE.
\end{acknowledgments}

\bibliography{mmag}




\end{document}



\title{Suplementary Material: Reflection of forward volume exchange spin waves at ferromagnetic junctions interpreted via continuity of angular momentum density flux}%


\author{Luis Sánchez-Tejerina}
 \email{luis.sanchez-tejerina@uva.es}
 \affiliation{Department of Electricity and Electronics, University of Valladolid, Valladolid, Spain.}%
\author{Luis M. Moreno-Ramírez}%
\affiliation{Department of Electricity and Electronics, University of Valladolid, Valladolid, Spain.}%
\author{Óscar Alejos Ducal}%
\affiliation{Department of Electricity and Electronics, University of Valladolid, Valladolid, Spain.}%

\date{\today}%


\maketitle


\section{Thin film approximation}

This study focuses on spin waves in the regime in which exchange interactions dominate over dipolar interactions. Therefore, we expect the magnetostatic field to play a minor role. We consider a thin slab for which the main demagnetizing tensor component is the one corresponding to the out-of-plane direction $z$, while the others are much smaller. Considering this and in order to simplify the analysis, we assumed that the solely effect of the dipolar field is to reduce the out-of-plane anisotropy by a factor proportional to the demagnetizing factor $N_z=0.9841$, neglecting the effect in the plane. This factor is derived from the Aharoni expressions~\cite{Aharoni:98} for the geometry of the ferromagnetic sample. This assumption corresponds to the thin film approximation adopted in the present work. To validate this approach, we compared micromagnetic results for two cases, i) neglecting the in-plane effects of the demagnetizing field and ii) computing the full demagnetizing kernel. In both cases, we excited the same geometry described in the main text with a single ferromagnetic material. We employed a $\mathrm{sinc}$ function to obtain the dispersion relations. The electric current considered through the heavy metal that causes the spin hall effect in this case is $J_{\mathrm{e}}\left(t\right)=J_0\,\mathrm{sinc}\left[\omega_{J,\mathrm{c}}\left( t-t_\mathrm{d} \right)\right]$, where the pulse amplitude is $J_0 = \qty{50}{\giga\ampere\per\meter\squared}$, the cut-off frequency $\omega_{J,\mathrm{c}}/\left(2\pi\right)=\qty{30}{\giga\hertz}$ and the delay time $t_{\mathrm{d}}=t_{\mathrm{s}}/2$ being $t_{\mathrm{s}}$ the total simulation time, set to $\qty{10}{\nano\second}$. This excitation configuration ensures that all frequencies below $\omega_{J,\mathrm{c}}$ are probed with a resolution of $\qty{100}{\mega\hertz}$.

\begin{figure*}
\centering
\includegraphics[width=0.85\textwidth]{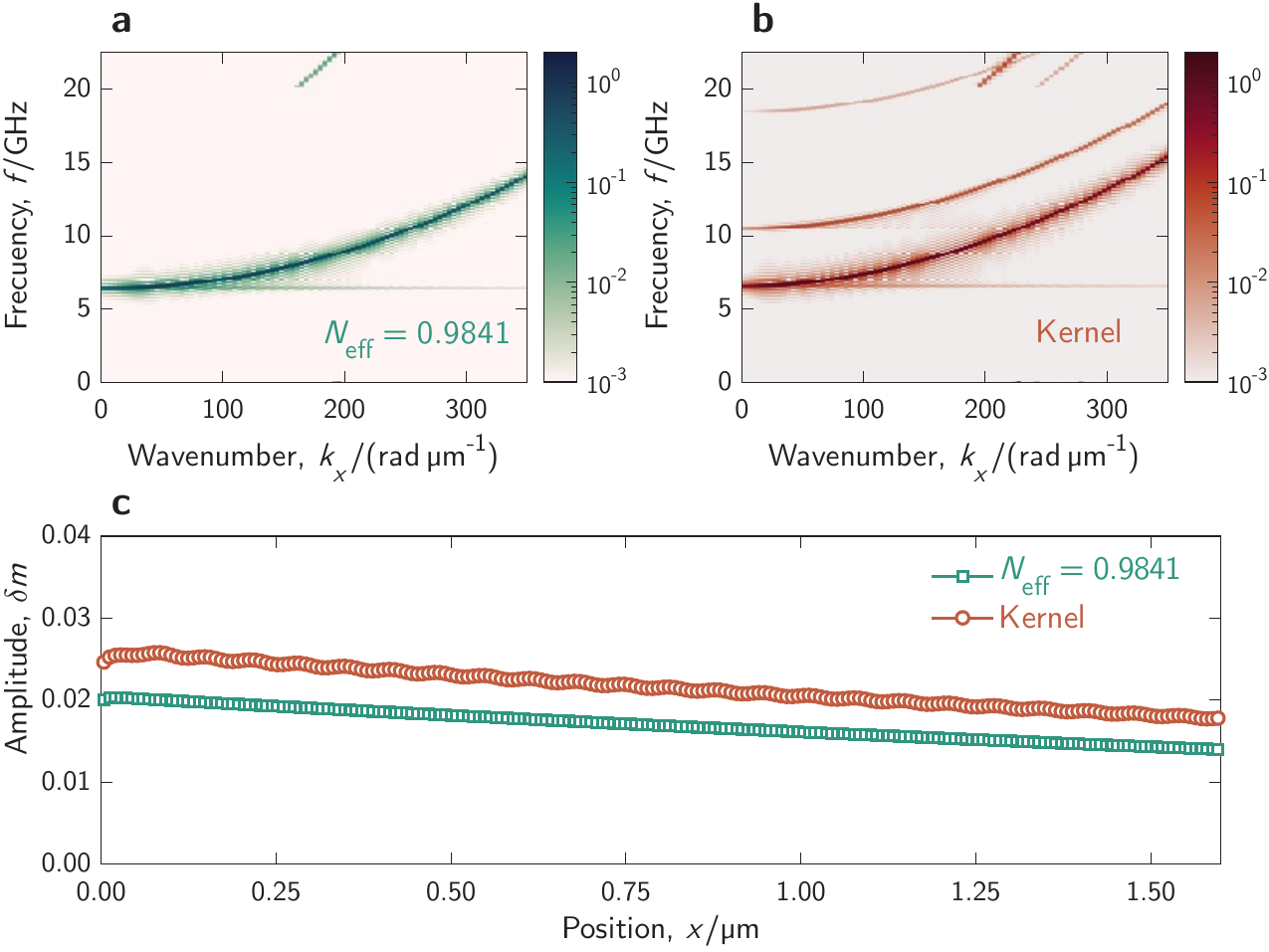}
\caption{\label{fig:S1} \textbf{Thin film approximation vs full demagnetizing kernel.} Dispersion relation derived from micromagnetic simulations (a) under the thin film approximation and (b) considering the demagnetizing kernel. (c) Spin wave amplitude as a function of the position along the strip relative to the excitation edge for both approaches. The plotted amplitude range is comparable to the amplitudes considered in this study.}

\end{figure*}

Fig.~\ref{fig:S1} depicts the micromagnetic results for (a) the simplified case under the thin film approximation and (b) the full micromagnetic simulations. Although panel (b) shows a richer spectrum, the characteristics of the first mode dependent on the wave vector $\vec{k}$, in which we are interested, are essentially identical.
Fig.~\ref{fig:S1}(c) shows the spin wave amplitude as a function of the distance from the excitation edge. In this case, the full micromagnetic simulations lead to a slightly larger amplitude as well as to a ripple in it. The latter characteristic is ascribed to the richer spectrum in this case, while the former is induced by the in-plane demagnetizing film. Despite the initial difference in amplitude, the amplitude decay is nearly the same. Given that the main characteristics of the first mode are nearly identical, we conclude that the thin film approximation is enough to carry out the study while keeping a clearer picture of the physics involved.

\section{Analytical calculations}
\subsection*{Analytical model}
As introduced in the main text, the medium supporting the propagation of SWs consists of a thin FM strip with out-of-plane uniaxial anisotropy defining the $z$ direction, with one in-plane direction much larger than the other, defining the $x$ direction. The excitation is produced uniformly in a reduced region along the $y$ and $z$ directions. Therefore, we consider propagation along the $x$ and no magnetization dependence along neither the $y$ nor $z$ directions. Additionally, we restrict the study to small amplitude excitation, thus disregarding excitation of frequencies different from the driver ($\omega_J$). We consider that the interactions relevant to our system are the exchange and anisotropy, so the total energy density is the sum of these two contributions $\varepsilon=\varepsilon_{\mathrm{exch}}+\varepsilon_{\mathrm{anis}}$. The exchange energy density $\varepsilon_{\mathrm{exch}} = A \left|\left| \nabla \vec{m} \right|\right|^2$, where $A$ is the exchange stiffness and $\left|\left| \nabla \vec{m} \right|\right|^2$ is the squared Frobenius norm of the normalized magnetization gradient tensor $\nabla \vec{m}$ and $\vec{m}=\vec{M}/M_{\mathrm{s}}$ the unit vector along the magnetization direction, $M_{\mathrm{s}}$ being the saturation magnetization and $\vec{M}$ the total magnetization. The anisotropy has two contributions, one from surface effects, $\varepsilon_{\mathrm{anis}} = K_{\mathrm{u}} \left(\vec{m}\hat{u}_z \right)^2$, and one from the demagnetizing field, $\varepsilon_{\mathrm{dmg}}=-\frac{1}{2}\vec{B}_{\mathrm{dmg}}\vec{M}$, where $\vec{B}_{\mathrm{dmg}}$ is the demagnetizing field depending on the demagnetizing factors~\cite{Aharoni:98}, (\textit{see Supplementary material S1}). The magnetization dynamics is given by the Landau-Lifshitz-Gilbert (LLG) equation~\cite{Sanchez-Tejerina:2025}

\begin{equation}\label{eq:LLG}
\frac{\mathrm{d}\vec{m}}{\mathrm{d}t} = -\gamma \vec{m}\times\vec{B}_{\mathrm{eff}}+\alpha \vec{m}\times\frac{\mathrm{d}\vec{m}}{\mathrm{d}t},
\end{equation}

\noindent where $\gamma$ is the gyromagnetic ratio, $\vec{B}_{\mathrm{eff}}$ is the effective field defined as the variational derivative of the total energy density, $\vec{B}_{\mathrm{eff}}=\frac{1}{M_{\mathrm{s}}}\frac{\delta \varepsilon}{\delta \vec{m}}$, and $\alpha$ is the Gilbert damping. With the above considered energy densities, the effective field can be written as follows

\begin{equation}\label{eq:eff1}
\vec{B}_{\mathrm{eff}}=\frac{2A}{M_{\mathrm{s}}}\nabla^2\vec{m}+\frac{2K_\mathrm{eff}}{M_{\mathrm{s}}}
m_{z}\hat{u}_{z},
\end{equation}

\noindent with $K_\mathrm{eff}=K_{\mathrm{u}}-\mu_{0} N_z M_{\mathrm{s}}^{2}/2$. Assuming small harmonic excitation and SW propagation along $x$, we can decompose the magnetization into equilibrium magnetization plus a small dynamic contribution 

\begin{equation}\label{eq:mag1}
\vec{m}=\vec{m}_{eq}+\delta \vec{m}e^{-i\left(\omega t - k_x x\right)}.
\end{equation}

\noindent Accordingly, we can write the effective field Eq.~\ref{eq:eff1} as an equilibrium field plus a dynamic contribution

\begin{equation}\label{eq:eff2}
\vec{B}_{\mathrm{eff}}=-\frac{2Ak_x^2}{M_{\mathrm{s}}}e^{-i\left(\omega t - k_x x\right)}\delta\vec{m}+\frac{2K_\mathrm{eff}}{M_{\mathrm{s}}}
m_{\mathrm{eq},z}\hat{u}_{z},
\end{equation}

\noindent where the exchange does not contribute to the equilibrium field and the anisotropy does not contribute to the dynamic field.

\subsection*{FVESWs progragation in a ferromagnet}
Since we are considering small excitations, then $m_{\mathrm{eq},z}\approx1$. Using Eq.~\ref{eq:mag1} and Eq.~\ref{eq:eff2} and disregarding second order terms\textcolor{LMR}{,} the LLG Eq.~\ref{eq:LLG} reads

\begin{equation}\label{eq:LLGsimpl}
i\omega S_{\mathrm{s}}\delta\vec{m} = \left( 2K_\mathrm{eff}
+ 2Ak_x^2 -i\omega S_{\mathrm{s}}\alpha \right) \left( \delta\vec{m} \times \hat{u}_{z} \right),
\end{equation}

\noindent where $S_{\mathrm{s}}=\frac{M_{\mathrm{s}}}{\gamma}$ is the saturation angular momentum density of the magnetic system. We can solve this equation by considering two mutually orthogonal linear polarizations or two circular polarizations; right-handed circular polarization (RHCP, $\frac{1}{\sqrt{2}} \left(\hat{u}_x+i\hat{u}_y\right)$) or left-handed circular polarization (LHCP, $\frac{1}{\sqrt{2}}\left(\hat{u}_x-i\hat{u}_y\right)$). These expressions are valid for positive frequencies, for the sign of the exponential argument chosen, and for the coordinate system considered (see Fig. 1 in the main text). It can be demonstrated that only RHCP can sustain a propagating wave. Thus, considering such an RHCP in \eqref{eq:LLGsimpl}, we have

\begin{equation}
i\omega S_{\mathrm{s}}\delta m \left( \hat{u}_x+i\hat{u}_y \right) = \delta m \left( 2K_\mathrm{eff}
+ 2Ak_x^2 -i\omega S_{\mathrm{s}}\alpha \right) \left( -\hat{u}_y + i\hat{u}_x \right) .
\end{equation}

\noindent For each cartesian component, we find the following

\begin{equation}
\omega S_{\mathrm{s}} = 2K_\mathrm{eff} + 2Ak_x^2 -i\omega S_{\mathrm{s}}\alpha ,
\end{equation}

\noindent and, therefore,

\begin{equation}
\omega = \frac{2K_\mathrm{eff} + 2Ak_x^2}{\left(1+i\alpha\right) S_{\mathrm{s}}} .
\end{equation}

\noindent This complex frequency can be expressed as $\omega=\omega'+i\omega''$ where $\omega'$ and $\omega''$ are real quantities

\begin{subequations}\label{eq:freq_ri}
\begin{align}
\omega'  &= \frac{2K_\mathrm{eff} + 2Ak_x^2}{\left(1+\alpha^2\right) S_{\mathrm{s}}},\\
\omega'' &=  -\alpha\frac{2K_\mathrm{eff} + 2Ak_x^2}{\left(1+\alpha^2\right) S_{\mathrm{s}}} .
\end{align}
\end{subequations}

\noindent It can be checked that the loss tangent $\omega''/\omega'$ is simply Gilbert damping. From Eq.~\ref{eq:freq_ri}(a) it is possible to derive the group velocity $\vec{v}_{\mathrm{gr}}$ as

\begin{equation}
\vec{v}_{\mathrm{gr}}=\frac{\partial\omega'}{\partial \vec{k}}=\frac{4Ak_x}{\left(1+\alpha^2\right) S_{\mathrm{s}}}\hat{u}_x.
\end{equation}

\noindent Additionally, $\omega''$ allows to compute the losses per unit time. A much more useful quantity would be the losses per unit length, which can be calculated as

\begin{equation}
\alpha_{\mathrm{gr}}=\frac{\omega''}{v_{\mathrm{gr}}}=-\alpha\frac{K_\mathrm{eff} + Ak_x^2}{2Ak_x}.
\end{equation}

\subsection*{Amplitude at the excitation region}
In the excitation region, the LLG equation~\ref{eq:LLG} is augmented by the torque due to the spin Hall effect (SHE)~\cite{Emori:14,Sanchez-Tejerina:20} 

\begin{equation}\label{eq:LLGaug}
\frac{\mathrm{d}\vec{m}}{\mathrm{d}t} = -\gamma \vec{m}\times\vec{B}_{\mathrm{eff}}+\alpha \vec{m}\times\frac{\mathrm{d}\vec{m}}{\mathrm{d}t}+\frac{\hbar \theta_{\mathrm{SH}}J_{\mathrm{e}}e^{-i\left(\omega_{J} t + \varphi\right)}}{2|e|d S_{\mathrm{s}}} \vec{m}\times\left(\vec{m} \times \vec{\sigma}_{\mathrm{SH}} \right),
\end{equation}

\noindent $\hbar$ being the reduced Planck constant, $\theta_{\mathrm{SH}}$ the spin Hall angle, $J_{e}$ the amplitude of the electric current density, $|e|$ the elementary charge, $\omega_J$ the angular frequency of the electric field, $\varphi$ the phase shift with respect to the dynamic magnetization, and $\vec{\sigma}_{\mathrm{SH}}$ the spin accumulation direction. To simplify the expression, we define the magnitude $\tau_{\mathrm{SH}}=\frac{\hbar \theta_{\mathrm{SH}}J_{\mathrm{e}}}{2|e|d}$. In the stationary state, damping and antidamping must compensate for each other. However, the torque given by the term $\alpha\vec{m}\times\frac{\mathrm{d}\vec{m}}{\mathrm{d}t}$ partially modifies the precession frequency (as given by Eq.~\ref{eq:freq_ri}(a)) and partially damps the system. 
To clearly distinguish how the trade off must be, it is better to write the LLG Eq.~\ref{eq:LLGaug} in the Landau-Lifshitz form
\begin{equation}\label{eq:LL}
\begin{split}
\left(1+\alpha^2\right)\frac{\mathrm{d}\vec{m}}{\mathrm{d}t} = &-\gamma \vec{m}\times\vec{B}_{\mathrm{eff}} + \tau_{\mathrm{SH}}e^{-i\left(\omega_{J} t + \varphi\right)}\vec{m}\times\left(\vec{m} \times \vec{\sigma}_{\mathrm{SH}} \right) +\\
&- \alpha \gamma \vec{m}\times\left(\vec{m}\times\vec{B}_{\mathrm{eff}}\right) - \alpha\tau_{\mathrm{SH}}e^{-i\left(\omega_{J} t + \varphi\right)}\vec{m} \times \vec{\sigma}_{\mathrm{SH}}.
\end{split}
\end{equation}

\noindent Thus, introducing the FVESW solutions and assuming electric current along $+y$ and spin current along $+z$, and consequently $\vec{\sigma}_{SH}=-\hat{u}_x$, the in-plane contributions read as follows

\begin{equation}
\begin{split}
-i\omega S_{\mathrm{s}}\left(1+\alpha^2\right)\delta m \left(\hat{u}_x + i\hat{u}_y \right) &= \delta m \left( 2K_\mathrm{eff} + 2Ak_x^2 \right) \left(\hat{u}_y -i\hat{u}_x \right) +\\
&+\sqrt{2}\tau_{\mathrm{SH}}e^{-i\left(\left(\omega_{J}-\omega \right)t + k_x x + \varphi\right)} \hat{u}_x +\\
&+\alpha \delta m \left( 2K_\mathrm{eff} + 2Ak_x^2 \right) \left( \hat{u}_x + i\hat{u}_y \right) +\\
&+ \sqrt{2}\alpha\tau_{\mathrm{SH}}e^{-i\left(\left(\omega_{J}-\omega \right)t + k_x x + \varphi\right)} \hat{u}_y.
\end{split}
\end{equation}

\noindent  As stated above, the damping and antidamping torque must be balanced and, therefore,

\begin{equation}
\sqrt{2}\tau_{\mathrm{SH}}e^{-i\left(\left(\omega_{J}-\omega \right)t + k_x x + \varphi\right)} \left( \hat{u}_x + \alpha\hat{u}_y \right) =-\alpha \delta m \left( 2K_\mathrm{eff} + 2Ak_x^2 \right) \left( \hat{u}_x + i\hat{u}_y \right) .
\end{equation}

\noindent Assuming $\omega_J=\omega$ and considering the excitation amplitude at the center of the excitation region $x=0$, $\delta m_{\mathrm{c}}$, for both $x$ and $y$ components:

\begin{subequations}
\begin{align}
\sqrt{2}\tau_{\mathrm{SH}}e^{-i\varphi} &=-\alpha \delta m_{\mathrm{c}} \left( 2K_\mathrm{eff} + 2Ak_x^2 \right) ,\\
\sqrt{2}\tau_{\mathrm{SH}}e^{-i\varphi} \alpha &=-i\alpha \delta m_{\mathrm{c}} \left( 2K_\mathrm{eff} + 2Ak_x^2 \right) .
\end{align}
\end{subequations}

\noindent Operating there

\begin{subequations}\label{eq:torque_equilibrium}
\begin{align}
\delta m_{\mathrm{c}}  &=\frac{-\tau_{\mathrm{SH}}\sqrt{2}}{\alpha\left( 2K_\mathrm{eff} + 2Ak_x^2 \right)} \cos{\varphi},\\
\delta m_{\mathrm{c}} &=\frac{\tau_{\mathrm{SH}}\sqrt{2}}{2K_\mathrm{eff} + 2Ak_x^2}\sin{\varphi} .
\end{align}
\end{subequations}

\noindent From Eqs.~\ref{eq:torque_equilibrium} we derived the phase shift

\begin{equation}
\varphi=\arctan{\frac{1}{-\alpha}} ,   
\end{equation}

\noindent (being $\sin\varphi>0$), and the amplitude in the center of the excitation region is

\begin{equation}
\delta m_{\mathrm{c}} = \frac{\tau_{\mathrm{SH}}\sqrt{2}}{2K_\mathrm{eff} + 2Ak_x^2}\sin{\varphi}.
\end{equation}

\noindent Therefore, the mean active angular momentum density injected at the center of the excitation region per cycle (assuming that the saturation angular momentum density of the FM does not depend on time) is

\begin{equation}
\frac{1}{T}\int_0^{T} S_{\mathrm{s}} \left( \delta m_{\mathrm{c}}  \right)^2 \cos^2\left({\omega t}\right) \mathrm{d}t = \frac{\tau_{\mathrm{SH}}}{2K_\mathrm{eff} + 2Ak_x^2}\sin{\varphi}.
\end{equation}

\noindent Moreover, the amplitude at the end of the excitation region ($\delta m_{\mathrm{e}}$) should consider the contributions of all points in the excitation region. All points on a surface perpendicular to the propagation direction contribute equally, but different surfaces contribute with different phases $e^{ik_x x}$. Summing all contributions in the entire excitation length $x_{\mathrm{e}}$, and dividing by the distance traveled per cycle $\Delta x= v_{\mathrm{gr}}/\omega$ leads to an additional factor that multiplies the stationary amplitude. This is simply averaging the angular momentum density that crossed a given surface normal to the propagation direction per cycle, but taking into account that only the excitation region is injecting signal to the sample. This factor is

\begin{equation}\label{eq:antenna_factor}
\frac{1}{\Delta x}\int_{-x_{\mathrm{e}}/2}^{x_{\mathrm{e}/2}} e^{i k_x x} \mathrm{d}x = \frac{\omega x_{\mathrm{e}}}{v_{\mathrm{gr}}}\mathrm{sinc}\left( k_x \frac{x_{\mathrm{e}}}{2}\right).
\end{equation}

\noindent Finally, it should be pointed out that the angular momentum injected into the system propagates in both directions $\pm x$. Therefore, the amplitude propagating towards $+x$ is halved. Taking into account these effects, the amplitude at the edge of the excitation region is

\begin{equation}
\delta m_{\mathrm{e}}=\frac{\omega x_{\mathrm{e}}}{4v_{\mathrm{gr}}}\frac{\tau_{\mathrm{SH}}\sin{\varphi}}{K_\mathrm{eff} + Ak_x^2}\mathrm{sinc}\left( k_x \frac{x_{\mathrm{e}}}{2}\right).
\end{equation}

\subsection*{Reflection and Transmission coefficients}
 The angular momentum density must be conserved at the interface between two different magnetic media, implying that, first, the direction of the angular momentum density is conserved and, second, the flux of the angular momentum density flux, defined as the magnitude times the velocity, $\vec{J}_{S,z}=S_{\mathrm{s}}\left(\delta m\right)^{2}\vec{v}_{\mathrm{gr}}$, is conserved. 
From the first condition, we get the following

\begin{equation}
\left(\delta \vec{m}^{\mathrm{i}}+\delta \vec{m}^{\mathrm{r}}\right)^2=\left(\delta \vec{m}^{\mathrm{t}}\right)^2 \Rightarrow \delta m^{\mathrm{i}}+\delta m^{\mathrm{r}}=\delta m^{\mathrm{t}} ,
\end{equation}

\noindent where $\delta \vec{m}^{\mathrm{i}}$, $\delta \vec{m}^{\mathrm{r}}$, and $\delta \vec{m}^{\mathrm{t}}$ are the incident, reflected, and transmitted amplitudes. From the second condition,

\begin{equation}
S_{\mathrm{s,1}}\left(\delta m^{\mathrm{i}}\right)^{2}v_{\mathrm{gr},1} - S_{\mathrm{s,1}}\left(\delta m^{\mathrm{r}}\right)^{2}v_{\mathrm{gr,1}} = S_{\mathrm{s,2}}\left(\delta m^{\mathrm{t}}\right)^{2}v_{\mathrm{gr,2}}.
\end{equation}

\noindent Defining the reflection and transmission coefficients as $R=\delta m^{\mathrm{r}}/\delta m^{\mathrm{i}}$ and $T=\delta m^{\mathrm{t}}/\delta m^{\mathrm{i}}$, they must fulfill

\begin{subequations}
\begin{align}
\frac{1-R}{1+R}=\frac{S_{\mathrm{s,2}}v_{\mathrm{gr,2}}}{S_{\mathrm{s,1}}v_{\mathrm{gr},1}},\label{eq:reflection}\\
T= \frac{2S_{\mathrm{s,1}}v_{\mathrm{gr},1}}{S_{\mathrm{s,1}}v_{\mathrm{gr},1}+S_{\mathrm{s,2}}v_{\mathrm{gr,2}}}\label{eq:transmission}.
\end{align}
\end{subequations}

\section{Excitation near the cut-off frequency}
\begin{figure*}
\centering
\includegraphics[width=0.85\textwidth]{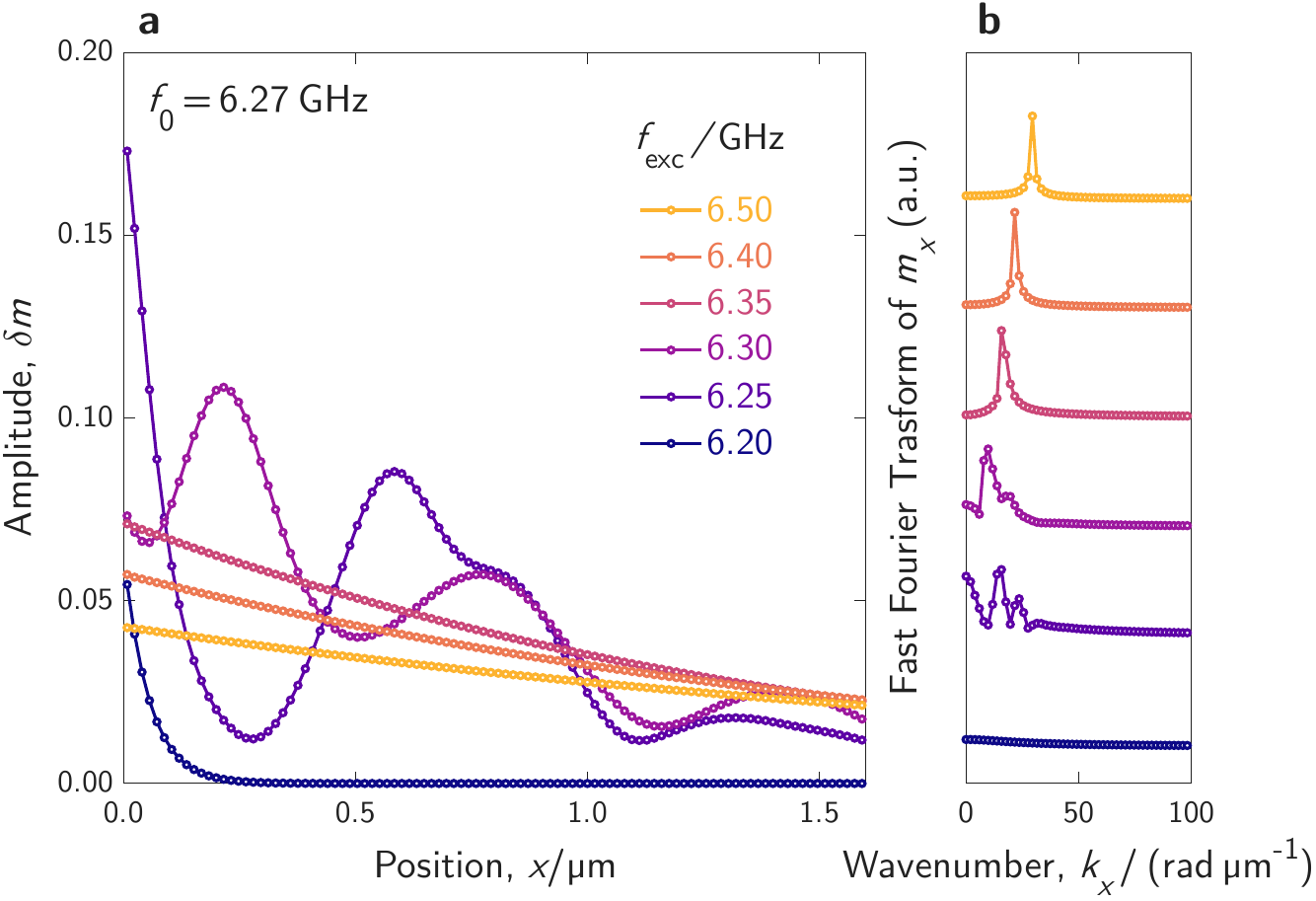}
\caption{\label{fig:S2} \textbf{FVESWs excitation near the cut-off frequency} (a) Spin wave amplitude as a function of the position along the strip relative to the excitation edge for different excitation frequencies near the cut-off frequency. (b) Fast Fourier transform of  the spin wave amplitude for different excitation frequencies near the cut-off frequency.}
\end{figure*}

The analytical model considers a single mode spin waveguide. Therefore, differences are expected in the cases where two or more modes are present. To assure comparability between analytical expressions and micromagnetic simulations, only the mode considered in the analytical derivation must be excited. Consequently, there is an upper limit to the excitation frequency which depends on the width and thickness of the ferromagnetic slab. Additionally, there is a lower limit to the applied frequency to excite spin waves. However, as the system approaches this cut-off frequency, the $k$-independent mode becomes excited, as it can be observed in Fig.~\ref{fig:S2}. A frequency sufficiently larger than the cut-off frequency should be used to avoid hybridization between the two modes. This effect can be observed in Fig.~\ref{fig:S2} where the characterized ferromagnetic thin slab with reference material parameters (\textit{see main text}) is excited for different frequencies. The material parameters impose a cut-off frequency of $f_0=\qty{6.27}{\giga\hertz}$, while the excitation frequencies range from $f=\qty{6.20}{\giga\hertz}$ to $f=\qty{6.50}{\giga\hertz}$. Fig.~\ref{fig:S2}(a) shows the spatial amplitude dependence of the spin wave amplitude along the strip, locating the origin at the edge of the excitation region. Fig.~\ref{fig:S2}(b) shows the Fourier transform of these amplitudes. At $f=\qty{6.20}{\giga\hertz}$ there is no excitation, as expected for frequencies clearly below the cut-off frequency. At $f=\qty{6.25}{\giga\hertz}\lesssim f_0$, there is signal transmitted to the spin waveguide, but with a wide spectrum rather than a well defined wavenumber. With increasing frequency the spectrum becomes thinner but still differs from a well defined harmonic signal. This trend is maintained up to $f \approx \qty{6.35}{\giga\hertz}$. At $f=\qty{6.50}{\giga\hertz}$ we find a well defined wavenumber, with an exponential decay along the strip with the same characteristics as derived from the analytical model. All the analyses performed in this study have used material parameters that ensure single mode excitation in all ferromagnetic materials, including cases involving joints.

\section{Reflection and transmission coefficients from stationary wave pattern fitting}
To obtain the reflection and transmission coefficients from the micromagnetic simulations, we fitted the spin wave stationary pattern on the first ferromagnetic medium and the exponential decay in the second medium, both depending on the reflection coefficient at the interface. The stationary wave pattern is obtained by adding the incident and reflected waves. For the incident wave, we have


\begin{equation}
\delta \vec{m^{\mathrm{i}}}(x,t)=\delta m^{\mathrm{i}} e^{\alpha_{\mathrm{gr,1}}x}e^{-i(\omega ' t-k_{x,1}x)}\frac{\left( \hat{u}_x+i\hat{u}_y \right)}{\sqrt{2}},
\end{equation}

\noindent propagating from the excitation region at $x=0$ to the interface between the two ferromagnetic materials at $x_{\mathrm{int}}$. The reflected wave can be expressed as a function of a reflection coefficient, $R$, which is generally complex. At the interface,


\begin{equation}
\delta m^{\mathrm{r}}(x_{\mathrm{int}},t) = R \delta m^{\mathrm{i}}(x_{\mathrm{int}},t).
\end{equation}

\noindent The reflected wave propagates within the same medium at the same frequency and, therefore, with the same spatial attenuation and wave number. Taking into account the change in propagation direction, the reflected wave is expressed as,


\begin{equation}
\delta \vec{m^{\mathrm{r}}}(x,t)=R\delta m^\mathrm{i} e^{\alpha_{\mathrm{gr,1}}\left(2x_{\mathrm{int}}-x\right)}e^{i2k_{x,1}x_{\mathrm{int}}}e^{-i(\omega ' t+k_{x,1}x)}\frac{\left( \hat{u}_x+i\hat{u}_y \right)}{\sqrt{2}}.
\end{equation}

\noindent The stationary wave in the first medium as a function of the amplitude of the incident wave at the interface, $\delta m^i\left(x_{\mathrm{int}}\right)=I$, would be


\begin{equation}
\delta \vec{m_1}(x,t)= I e^{\alpha_{\mathrm{gr,1}}\left(x-x_{\mathrm{int}}\right)}e^{-i(\omega ' t-k_{x,1}x)}\frac{\left( \hat{u}_x+i\hat{u}_y \right)}{\sqrt{2}}\left(1 + R e^{2\alpha_{\mathrm{gr,1}}\left(x_{\mathrm{int}}-x\right)}e^{i2k_{x,1} \left(x_{\mathrm{int}}-x\right)}\right),
\end{equation}

\noindent and its magnitude, defining $\bar{x}=x-x_{\mathrm{int}}$,


\begin{equation}
\delta m_1(x,t) = \sqrt{\delta \vec{m_1}^{*}(x,t)\cdot \delta \vec{m_1}(x,t)}= I \sqrt{e^{2\alpha_{\mathrm{gr,1}}\bar{x}} + |R|^2 e^{-2\alpha_{\mathrm{gr,1}}\bar{x}} + 2|R| \cos{\left(2k_{x,1} \bar{x} - \phi\right)}},
\end{equation}
where $|R|$ and $\phi$ are the modulus and phase of the reflection coefficient, respectively. Within the second FM medium, only the transmitted wave propagates. The transmitted amplitude can be expressed as a function of the incident amplitude via the transmission coefficient


\begin{equation}
\delta \vec{m}_2(x,t) = T I e^{\alpha_{\mathrm{gr},2}\left(x-x_{\mathrm{int}}\right)}e^{-i(\omega ' t-k_{x,2}x)}\frac{\left( \hat{u}_x+i\hat{u}_y \right)}{\sqrt{2}},
\end{equation}

\noindent and its magnitude,


\begin{equation}
\delta m_2(x,t) = I e^{\alpha_{\mathrm{gr,2}}\bar{x}} \sqrt{T^*T} = I e^{\alpha_{\mathrm{gr,2}}\bar{x}} \sqrt{1 + |R|^2 + 2|R| \cos{\phi}},
\end{equation}

\noindent where it has been taken into account that $T=1+R$.

\begin{acknowledgments}
We acknowledge funding from Project {PID2023-150853NB-C31} funded by MICIU/AEI /10.13039/501100011033 and by European Union Next GenerationEU/PRTR.
\end{acknowledgments}

\bibliography{mmag}


